\documentclass[conference]{IEEEtran}
\IEEEoverridecommandlockouts
\usepackage{cite}
\usepackage{amsmath,amssymb,amsfonts}
\usepackage{algorithmic}
\usepackage{graphicx}
\usepackage{textcomp}
\usepackage{xcolor}
\usepackage{tikz}
\usepackage{pgfplots}
\pgfplotsset{compat=1.17}
\usetikzlibrary{positioning, arrows.meta, shapes.geometric}
\usepackage{booktabs}
\usepackage{multirow}
\usepackage{array}

\def\BibTeX{{\rm B\kern-.05em{\sc i\kern-.025em b}\kern-.08em
    T\kern-.1667em\lower.7ex\hbox{E}\kern-.125emX}}

\begin{document}

\title{Arapai: An Offline Adaptive AI Tutoring System for\\
Competency-Based Learning in Low-Connectivity Environments}

\author{
\IEEEauthorblockN{Joseph Walusimbi, Ann Move Oguti, Joshua Benjamin Ssentongo, and Keith Ainebyona}
\IEEEauthorblockA{
\textit{Department of Electronics and Computer Engineering, Soroti University}\\
Soroti, Uganda\\
2401600068@sun.ac.ug, amoguti@sun.ac.ug, 2401600236@sun.ac.ug, 2401600230@sun.ac.ug}
}

\maketitle

\begin{abstract}
Persistent learning deficits in low-resource educational environments stem from a structural mismatch: competency-based curricula demand individualized, iterative feedback, yet most digital learning tools require stable internet connectivity that is unavailable to millions of students. This paper presents Arapai, a fully offline adaptive AI tutoring system designed to close this gap through on-device intelligence. Arapai integrates a quantized lightweight language model with a four-stage pedagogical loop (diagnosis, instruction, worked example, and adaptive feedback) implemented without any external infrastructure. Following development, Arapai was demonstrated to a broad cross-section of stakeholders across multiple national academic and institutional exhibitions in Uganda, including technical judges, university faculty, tertiary educators, and secondary school practitioners. Expert observers provided informal feedback affirming Arapai's relevance and contextual fit, particularly with respect to Uganda's newly enacted Competency-Based Curriculum (CBC). These demonstrations establish a foundation of expert-observed validation; rigorous controlled efficacy studies with student cohorts are planned as the immediate next research phase. Arapai demonstrates that structured offline AI can function as viable educational infrastructure in low-connectivity emerging-market contexts, offering a practical pathway to AI-augmented learning without internet dependence.
\end{abstract}


\section{Introduction}

Despite decades of investment in educational technology, measurable learning gains in low-resource settings remain elusive. A 2020 World Bank report found that more than half of children in low- and middle-income countries cannot read a simple passage by age~10, a phenomenon termed \textit{learning poverty} \cite{b5}. In Sub-Saharan Africa, these outcomes are compounded by high student-to-teacher ratios, inconsistent teacher deployment, and near-absent mechanisms for individualized feedback \cite{b14}.

The shift toward competency-based curricula (CBC), adopted in Uganda through the Ministry of Education and Sports framework \cite{b4} exacerbates this tension. CBC explicitly requires students to demonstrate applied understanding through problem-solving and transfer tasks, not mere recall. Achieving this demands sustained, corrective feedback at the individual level: a resource that classroom instruction at scale cannot reliably provide.

This challenge carries particular urgency in Uganda's national development context. Uganda Vision 2040, the country's long-term blueprint for transforming into a modern, prosperous middle-income country, explicitly frames education as a fundamental human right and positions investment in human capital as central to realizing its aspirations \cite{b16}. The Fourth National Development Plan (NDP IV, 2025/26--2029/30) reinforces this commitment, prioritizing human capital development and the expansion of quality education as a prerequisite for sustainable socio-economic transformation \cite{b17}. These ambitions are further anchored to global development imperatives: SDG~4 calls for inclusive and equitable quality education and lifelong learning opportunities for all, while SDG~10 demands the reduction of inequality within and among countries---both goals that remain structurally out of reach when educational tools are gated behind connectivity. Yet the infrastructure reality on the ground stands in sharp tension with these ambitions. Despite a national internet penetration rate of approximately 27\% as of 2024, only 9\% of rural Ugandans access the internet \cite{b18}, and more than 72\% of the population lives in rural areas \cite{b19}. Rural schools are disproportionately unconnected, with the digital divide closely mirroring Uganda's geography \cite{b20}. Even in areas where connectivity nominally exists, the recurring cost of mobile data bundles places consistent internet access out of reach for the majority of students and teachers. This means that online learning tools, however well-designed, are structurally inaccessible to precisely the learners that Vision 2040 and NDP IV most urgently seek to reach.

Intelligent tutoring systems (ITS) have repeatedly demonstrated that personalized feedback can substitute for, and in some studies surpass, one-on-one human tutoring \cite{b1, b11}. However, virtually every deployed ITS, from Khan Academy to Carnegie Learning, assumes persistent broadband connectivity, rendering these platforms structurally inaccessible in rural Uganda and analogous environments \cite{b3, b13}.

Recent advances in compact language model architectures, including quantized sub-3B parameter models, have made high-quality on-device inference feasible on commodity hardware \cite{b9, b10}. Yet no prior work has integrated these models into a full pedagogical pipeline aligned to a defined CBC context and subjected to structured expert review across diverse educational stakeholders.

This paper makes the following contributions:
\begin{itemize}
    \item A complete system architecture for offline AI-assisted tutoring, designed specifically for CBC-aligned secondary education in low-connectivity environments.
    \item A hybrid instructional engine that combines generative AI explanations with deterministic rule-based response evaluation, ensuring pedagogical reliability without internet dependence.
    \item A multi-event stakeholder validation study spanning technical judges, university faculty, tertiary institutions, and secondary school educators at nationally prominent exhibitions and innovation forums in Uganda.
    \item A deployment model targeting schools, NGOs, and government digital-education programs, with a pathway to mobile optimization.
\end{itemize}

\section{Related Work}

\subsection{Intelligent Tutoring Systems}
Anderson et al.\ established the foundational evidence that model-tracing tutors providing immediate corrective feedback can double the learning rate compared to classroom instruction \cite{b1}. D'Mello and Graesser extended this paradigm to affective and conversational tutoring, demonstrating that responsiveness to student confusion and frustration further amplifies outcomes \cite{b11}. A systematic review by Sahu et al.\ confirms that AI-driven educational applications consistently outperform passive digital content across K-12 and undergraduate cohorts \cite{b12}.

\subsection{Connectivity Barriers in Digital Education}
UNESCO's global AI-in-education policy brief identifies connectivity as the primary barrier to ITS adoption in low-income countries \cite{b3}. The World Bank's 2023 technology-in-education report notes that infrastructure-dependent tools entrench inequality rather than address it \cite{b13}. Unwin's longitudinal analysis of digital education in Africa further emphasizes that tools designed for high-connectivity contexts fail to transfer to African classroom realities \cite{b14}.

\subsection{Compact and On-Device Language Models}
Google DeepMind's Gemma family \cite{b9} and Microsoft Research's Phi-2 \cite{b10} demonstrate that models with fewer than 3 billion parameters, when quantized to 4-bit precision, can perform curriculum-level explanation and question-answering tasks on CPU-only hardware. However, these models have not been embedded within structured pedagogical frameworks or subjected to expert review within CBC-specific educational contexts.

\subsection{Prior Offline Educational Software}
Several systems have previously targeted low-connectivity educational environments and constitute important prior art. KA Lite, the offline distribution of Khan Academy, demonstrated that pre-packaged video content can be deployed via Raspberry Pi servers in schools without internet access, but it provides no adaptive feedback or interactive response evaluation \cite{b21}. The Remote Area Community Hotspot for Education and Learning (RACHEL) similarly packages curated static content for offline access without learner modeling or personalized progression \cite{b22}. Kolibri (formerly KA Lite) extended this paradigm with basic exercise sets but still relies on fixed question banks without generative explanation capability \cite{b23}. The present system advances beyond these platforms by integrating a generative language model for dynamic explanation, rule-based adaptive evaluation, and a longitudinal learner model, all without any internet dependency.

\subsection{Gap Addressed}
The intersection of offline-capable language models, CBC-aligned pedagogy, and structured expert validation in Sub-Saharan educational contexts remains underexplored. Prior offline education platforms (KA Lite, RACHEL, Kolibri) provide static content without adaptive AI feedback; prior ITS research assumes connectivity. This work addresses that gap directly, presenting both the design of Arapai and an account of its reception across a diverse, nationally distributed stakeholder base.

\section{System Architecture}

Arapai is designed around three principles: \textit{reliability} (deterministic behavior in adversarial connectivity conditions), \textit{pedagogical fidelity} (alignment with CBC instructional sequences), and \textit{hardware accessibility} (operation on devices costing under USD 200).

\begin{figure}[tbp]
\centering
\begin{tikzpicture}[x=1cm, y=1cm]
  \tikzstyle{box}  = [rectangle, draw=black, line width=0.6pt,
                      align=center, minimum height=0.72cm, font=\small]
  \tikzstyle{sbox} = [rectangle, draw=black, line width=0.6pt,
                      align=center, minimum height=0.90cm,
                      font=\footnotesize, fill=gray!15]
  \tikzstyle{arr}  = [->, >=stealth, line width=0.6pt]

  \node[box, minimum width=2.8cm] (student) at (3.2, 5.2)
        {Student (User)};
  \node[box, minimum width=4.8cm] (ui) at (3.2, 4.0)
        {User Interface (Mobile / PC)};
  \node[box, minimum width=6.6cm, minimum height=1.0cm, fill=gray!10] (engine) at (3.2, 2.7)
        {\textbf{Learning Engine}\\
         \scriptsize Diagnosis $\to$ Instruction $\to$ Practice $\to$ Feedback};
  \node[sbox, minimum width=2.0cm] (llm) at (0.8, 1.0)
        {On-Device AI\\ \scriptsize (Quantized LLM)};
  \node[sbox, minimum width=2.0cm] (kb)  at (3.2, 1.0)
        {Curriculum KB\\ \scriptsize (CBC Content)};
  \node[sbox, minimum width=2.0cm] (db)  at (5.6, 1.0)
        {Local Storage\\ \scriptsize (Progress / Scores)};

  \draw[arr] (student.south) -- (ui.north);
  \draw[arr] (ui.south) -- (engine.north);

  \coordinate (busC) at (3.2, 1.85);
  \coordinate (busL) at (0.8, 1.85);
  \coordinate (busR) at (5.6, 1.85);

  \draw[arr] (engine.south) -- (busC);
  \draw      (busL) -- (busR);
  \draw[arr] (busL) -- (llm.north);
  \draw[arr] (busC) -- (kb.north);
  \draw[arr] (busR) -- (db.north);
\end{tikzpicture}
\caption{System architecture. All components reside on-device; no external network calls are required at runtime.}
\label{fig:arch}
\end{figure}

As shown in Fig.~\ref{fig:arch}, Arapai comprises five integrated components:

\textbf{User Interface.} A Streamlit-based front-end provides a consistent cross-platform experience on both PC and Android. The interface surfaces diagnostic questions, explanations, practice items, and feedback within a unified session view.

\textbf{Learning Engine.} The engine is Arapai's orchestration layer. It manages session state, selects and sequences instructional content from the knowledge base, invokes the language model for explanation generation, applies rule-based evaluation to student responses, and updates the local progress database. Critically, all branching logic is deterministic, ensuring predictable behavior independent of model stochasticity.

\textbf{On-Device AI Model.} Arapai uses Gemma 2B \cite{b9} quantized to INT4 precision via GGUF format. Quantization reduces memory footprint from approximately 4~GB (FP16) to under 1.2~GB, enabling inference on devices with 4~GB RAM without a GPU. The model is invoked exclusively for \textit{generation} tasks (explanations, worked examples); it does not evaluate student answers.

\textbf{Curriculum Knowledge Base.} A structured JSON repository of topics, subtopics, learning objectives, example problems, and worked solutions aligned to Uganda's O-level CBC mathematics scope and sequence \cite{b4}.

\textbf{Local Progress Database.} An SQLite database persists per-student diagnostic scores, attempt histories, mastery flags, and session logs. This enables the adaptive engine to maintain longitudinal learner models across sessions without any cloud synchronization.

\section{Adaptive Learning Framework}

The instructional design is grounded in Bloom's Taxonomy of Educational Objectives \cite{b15} and the corrective-feedback literature \cite{b1, b11}. Bloom's Taxonomy organizes cognitive learning into six hierarchical levels: \textit{Remember}, \textit{Understand}, \textit{Apply}, \textit{Analyze}, \textit{Evaluate}, and \textit{Create}, and holds that durable competency requires deliberate progression through these levels rather than exposure to content alone. This principle maps directly onto the demands of Uganda's CBC framework \cite{b4}, which explicitly requires students to demonstrate applied and analytical understanding, not mere recall.

Table~\ref{tab:bloom} makes the mapping between each system stage and Bloom's levels concrete. The \textit{Diagnosis} stage targets \textit{Remember} and \textit{Understand} by asking students to recall definitions and identify correct factual relationships (e.g., ``State the condition under which a quadratic has two distinct real roots''). The \textit{Instruction} stage targets \textit{Understand} by generating tailored conceptual explanations that build relational knowledge. The \textit{Worked Example} stage bridges \textit{Understand} and \textit{Apply} by modeling step-by-step procedural reasoning in context (e.g., demonstrating completing-the-square on a specific equation). The \textit{Practice and Feedback} stage engages \textit{Apply} and \textit{Analyze} by requiring students to solve novel problems and receive error-specific corrective feedback that names and explains the precise misconception.

\begin{table}[tbp]
\caption{Mapping of System Stages to Bloom's Taxonomy Levels}
\label{tab:bloom}
\begin{center}
\begin{tabular}{p{2.2cm} p{1.9cm} p{3.3cm}}
\toprule
\textbf{System Stage} & \textbf{Bloom's Level(s)} & \textbf{Example Item (Quadratics)} \\
\midrule
Diagnosis       & Remember, Understand & ``How many solutions does $x^2 - 4 = 0$ have?'' \\
Instruction     & Understand           & LLM explanation of the discriminant \\
Worked Example  & Understand, Apply    & Step-by-step: solve $x^2+5x+6=0$ by factoring \\
Practice \& Feedback & Apply, Analyze  & ``Solve $2x^2 - 3x - 2 = 0$ using the quadratic formula'' \\
\bottomrule
\end{tabular}
\end{center}
\end{table}

\begin{figure}[tbp]
\centering
\includegraphics[width=\columnwidth]{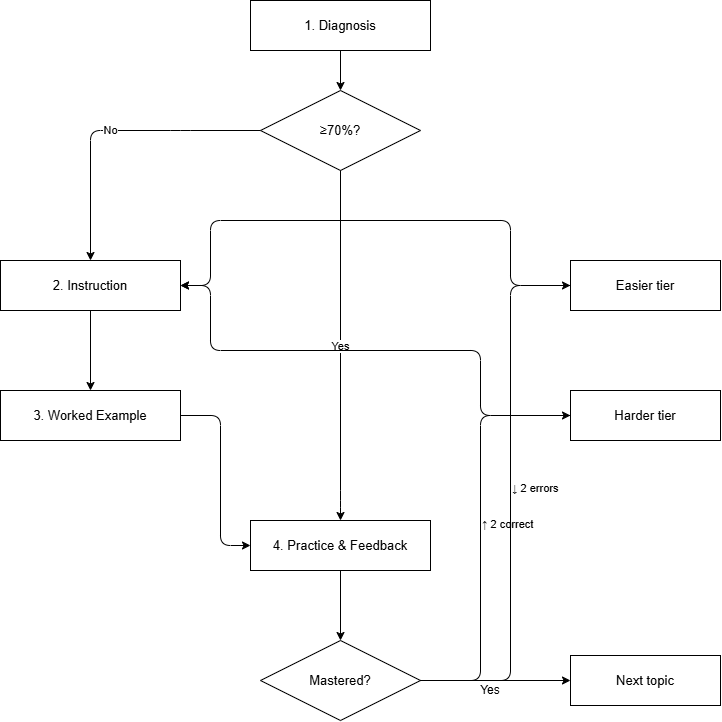}
\caption{Adaptive learning loop. Difficulty tier is adjusted after each
practice attempt; students returning below mastery re-enter at the
instruction or diagnosis stage.}
\label{fig:loop}
\end{figure}

The adaptive difficulty controller ensures that as mastery is achieved at one level, challenge is escalated toward the next, embodying Bloom's foundational insight that learning is cumulative and must be scaffolded. This explicit cognitive architecture distinguishes Arapai from passive content-delivery tools and positions it as a principled pedagogical instrument aligned with both established learning theory and the competency-demonstration philosophy of the CBC. Each learning session follows this four-stage loop, illustrated in Fig.~\ref{fig:loop}.

\textbf{Stage 1: Diagnosis.} The engine presents a short set of probing questions to assess prior knowledge on the target subtopic. Responses are evaluated by the rule-based scorer; a mastery threshold of 70\% determines whether the student proceeds directly to practice or enters the instructional sequence.

\textbf{Stage 2: Instruction.} For students below mastery, the language model generates a concise conceptual explanation ($\leq$150 tokens) tailored to the difficulty level flagged by diagnosis. Explanations are generated via a structured prompt template that encodes three parameters: the subtopic name, the identified difficulty tier (foundational, intermediate, or advanced), and the specific misconception pattern detected by the diagnostic scorer. Arapai prompt instructs the model to act as a secondary-school mathematics tutor and to produce a plain-language explanation of no more than 150 tokens, avoiding symbolic manipulation in favor of conceptual framing. An example prompt structure is shown below:

\begin{quote}
\texttt{[SYSTEM] You are a secondary-school mathematics tutor. Explain the following concept clearly and concisely in 150 words or fewer. Do not solve an equation; explain the concept only.}\\
\texttt{[USER] Topic: Quadratic equations. Subtopic: Discriminant. Difficulty: foundational. Student error pattern: confused sign of discriminant with number of roots. Provide a corrective conceptual explanation.}
\end{quote}

The difficulty tier is passed as an explicit token in the user turn, and the error-pattern field is populated by the rule-based evaluator's classification of the student's most recent incorrect response. This templating strategy constrains the model's output space, maintaining factual correctness while leveraging its natural language generation capability.

\textbf{Stage 3: Worked Example.} The engine retrieves a solved problem from the knowledge base corresponding to the student's identified weakness. Worked examples are structured in step-annotated format to scaffold procedural understanding.

\textbf{Stage 4: Practice and Feedback.} The student attempts a new problem. The rule-based evaluator parses and checks the response, then the language model generates targeted corrective feedback referencing the specific error type. The adaptive controller increments difficulty after two consecutive correct responses and decrements after two consecutive errors.

\subsection{Design Rationale: Hybrid Rule-Based and Generative Architecture}

The decision to use deterministic rule-based evaluation rather than LLM-based evaluation for student response checking is the most consequential architectural decision in Arapai. A compressed 2B-parameter model quantized to INT4 precision operates at a significant capacity disadvantage relative to full-scale frontier models. When asked to judge the mathematical correctness of a student's algebraic steps (a task that requires symbolic manipulation verification, not natural language generation), the model is likely to hallucinate confirmations or miss structural errors. The consequences of false-positive evaluation in a tutoring context are pedagogically severe: a student told their incorrect solution is correct will reinforce the misconception, precisely the opposite of Arapai's purpose.

The rule-based evaluator avoids this failure mode entirely. For the O-level mathematics domain covered by the knowledge base, correct answers and valid solution paths can be enumerated and checked algorithmically with 100\% accuracy. The LLM is then reserved exclusively for the tasks where generative capability is genuinely advantageous: producing natural-language conceptual explanations and formulating corrective feedback in terms the student can understand. This division of labor (generative AI for fluency, deterministic logic for correctness) is not a workaround but a principled design choice that exploits the comparative advantages of each component. Future work will investigate whether larger on-device models can reliably perform evaluation on open-ended question types, but for the current scope this hybrid approach provides both reliability and educational integrity.

\section{Implementation}

\subsection{Technology Stack}
The prototype is implemented in Python~3.11. The front-end uses Streamlit~1.32; the inference back-end uses \texttt{llama-cpp-python} to load and run the GGUF-quantized Gemma 2B model. The curriculum knowledge base and student progress are stored in SQLite~3. The full application, including model weights, is packaged as a self-contained installer ($\approx$1.5~GB) requiring no internet connection after initial distribution.

\subsection{Quantization and Inference Optimization}
Gemma 2B was converted to INT4 GGUF format, reducing peak VRAM/RAM usage to 1.1~GB. On a mid-range laptop (Intel Core i5-8th gen, 8~GB RAM, no discrete GPU), mean inference latency for a 150-token explanation is 3.2~seconds (measured over 50 independent inference calls), well within an acceptable interactive threshold. Context windows are capped at 512 tokens per turn to maintain consistent response times on lower-specification hardware.

The rule-based evaluator achieves 100\% accuracy on the closed-answer item types in the current knowledge base (multiple-choice, numeric answer, and single-step algebraic expressions), as correct answers are enumerated exhaustively and checked by string normalization and symbolic equivalence. Coverage of the current knowledge base relative to the full O-level mathematics syllabus is deliberately narrow: the prototype covers one of approximately 14 O-level mathematics units. Table~\ref{tab:benchmarks} summarizes the key performance characteristics of the current prototype.

\begin{table}[tbp]
\caption{System Performance Benchmarks (Prototype)}
\label{tab:benchmarks}
\begin{center}
\begin{tabular}{p{3.8cm} p{3.8cm}}
\toprule
\textbf{Metric} & \textbf{Value} \\
\midrule
Mean inference latency (150 tokens) & 3.2 s (i5-8th gen, 8 GB RAM, no GPU) \\
Model RAM footprint (INT4 GGUF)     & 1.1 GB \\
Rule-based evaluator accuracy       & 100\% on enumerable answer types \\
Knowledge base: topics covered      & 1 of $\approx$14 O-level math units \\
Knowledge base: practice items      & 90 (6 subtopics $\times$ 3 tiers $\times$ 5 items) \\
Installer size                      & $\approx$1.5 GB \\
\bottomrule
\end{tabular}
\end{center}
\end{table}

\subsection{Curriculum Alignment}
The knowledge base for this prototype covers the Uganda CBC O-level mathematics unit on quadratic equations, comprising six subtopics: factorization, completing the square, the quadratic formula, discriminant analysis, word problems, and graphical interpretation. Each subtopic contains three difficulty tiers (foundational, intermediate, advanced) with five practice items and three worked examples per tier, yielding 90 practice items in total.

\section{Stakeholder Evaluation and Expert Reception}

\subsection{Evaluation Approach}

Rather than a formal controlled trial (which constitutes the next phase of this work), Arapai was evaluated through expert demonstration and informal feedback collection across a series of national exhibitions and innovation forums in Uganda. Participants interacted with live demonstrations of Arapai, engaged with its instructional workflows, and provided verbal and written informal feedback on its features, usability, and applicability to CBC delivery. No structured questionnaires, Likert-scale instruments, or quantitative scoring rubrics were administered at these events; the evidence reported here is qualitative in nature, drawn from observer commentary and informal discussion. This approach was selected pragmatically, given the exhibition context, and is acknowledged as a limitation. The findings should be read as expert-informed practitioner feedback, not as quantitative usability or efficacy data.

Reviewers encompassed a deliberately broad stakeholder spectrum: technical judges with expertise in software systems and AI, university faculty from multiple institutions, educators from tertiary institutions, and secondary school practitioners directly engaged with CBC delivery.

\subsection{Exhibition and Event Contexts}

Arapai was presented at four nationally significant venues, summarized in Table~\ref{tab:events}.

\begin{table}[tbp]
\caption{Summary of Stakeholder Exposure Events}
\label{tab:events}
\begin{center}
\begin{tabular}{p{2.2cm} p{1.0cm} p{1.5cm} p{1.3cm} p{1.2cm}}
\toprule
\textbf{Event} & \textbf{Date} & \textbf{Reviewer Profile} & \textbf{Approx. Reviewers} & \textbf{Feedback Type} \\
\midrule
Gulu NCHE Exhibition 2026 & Jan.\ 2026 & Technical judges, higher education assessors, students & $\approx$15 & Informal verbal \\
Mbarara University Innovation Week 2026 & Feb.\ 2026 & University faculty, researchers, student innovators & $\approx$20 & Informal verbal \\
Launch of 40th TISC at Soroti University (URSB) & Mar.\ 2026 & National IP policymakers, institutional leaders, technology incubators & $\approx$30 & Informal verbal \\
Secondary \& Tertiary Institution Engagements & Ongoing & Secondary school teachers, CBC practitioners, tertiary educators & $\approx$25 & Informal verbal \& written \\
\bottomrule
\end{tabular}
\end{center}
\end{table}

\textbf{Gulu NCHE Exhibition 2026.} Arapai was exhibited at a National Council for Higher Education (NCHE) event in Gulu, where it was reviewed by technical judges and higher education assessors responsible for evaluating innovation quality and institutional relevance. Judges assessed Arapai against criteria that include technical soundness, originality, and applicability to Uganda's educational landscape.

\textbf{Mbarara University Innovation Week 2026.} At Mbarara University of Science and Technology's Innovation Week, Arapai was presented to a mixed audience of university faculty, applied researchers, and student innovators. This setting provided rigorous peer-level technical scrutiny alongside pedagogical commentary from academics directly involved in teacher training and curriculum design.

\textbf{Launch of the 40th Technology Support and Incubation Centre (TISC) at Soroti University.} Organized by the Uganda Registration Services Bureau (URSB), the launch of the 40th TISC represented one of the most prominent institutional showcases in which Arapai was featured. Attendees included national policymakers, intellectual property authorities, and technology incubation leaders, a constituency uniquely positioned to assess Arapai's scalability, replicability, and policy alignment. Reviewers at this event appreciated Arapai's potential impact and recognized its relevance to Uganda's national technology and education landscape.

\textbf{Secondary and Tertiary Institution Engagements.} Beyond formal exhibitions, Arapai was demonstrated to educators and practitioners from secondary schools and tertiary institutions who interact daily with CBC delivery challenges. This ground-level exposure generated the most contextually grounded feedback, with reviewers affirming Arapai's direct applicability to classroom realities they encounter firsthand.

\subsection{Reviewer Feedback and Findings}

Across all events and reviewer categories, the reception was consistently and strongly positive. Key themes that emerged from reviewer feedback are as follows.

\textbf{CBC Alignment.} Reviewers with direct familiarity with Uganda's newly enrolled CBC framework uniformly affirmed that Arapai's instructional design, specifically its diagnostic-first approach, worked-example scaffolding, and adaptive feedback, maps closely onto the competency demonstration requirements of the curriculum. Several secondary school educators noted that Arapai addresses a gap they experience acutely in classroom practice: the inability to provide individualized corrective feedback at scale.

\textbf{Offline Capability as a First-Class Feature.} Technical judges and institutional reviewers consistently highlighted the fully offline architecture as a distinguishing and practically critical innovation. In a landscape where most edtech tools are rendered non-functional by the intermittent connectivity that characterizes Ugandan schools, Arapai's ability to deliver AI-quality instruction entirely on-device was described as genuinely differentiated.

\textbf{Usability and Accessibility.} Practitioners from secondary and tertiary institutions praised Arapai's interface for its clarity and low barrier to entry, noting that it does not presuppose familiarity with AI tools or advanced computing environments. The consistent interactive experience across the diagnostic, instructional, and feedback stages was recognized as pedagogically coherent.

\textbf{Institutional and Policy Relevance.} Reviewers at the URSB TISC launch and NCHE exhibition situated Arapai within Uganda's broader digital transformation agenda, noting its alignment with national priorities for technology-driven education reform. Arapai's design as a complement to, rather than a replacement of classroom teachers was particularly well-received by policy-oriented stakeholders.

\subsection{Summary of Expert Validation}

The breadth and consistency of positive expert observer feedback across technically rigorous, institutionally prominent, and pedagogically grounded contexts constitutes meaningful practitioner-level validation of Arapai's design and relevance. Table~\ref{tab:feedback} summarizes the primary dimensions on which observers commented and the nature of feedback received. These findings are qualitative and reflect informal expert observation; they do not constitute quantitative evidence of efficacy.

\begin{table}[tbp]
\caption{Summary of Expert Observer Informal Feedback by Dimension}
\label{tab:feedback}
\begin{center}
\begin{tabular}{p{3.2cm} p{4.4cm}}
\toprule
\textbf{Dimension} & \textbf{Observer Feedback (Informal)} \\
\midrule
CBC curriculum alignment      & Affirmed by secondary and tertiary educators with direct CBC delivery experience \\
Offline architecture value    & Highlighted as critical practical differentiator by technical judges and institutional reviewers \\
Technical soundness           & Positively received by NCHE judges and university faculty \\
Usability                     & Noted as accessible across all practitioner groups \\
Policy and institutional fit  & Recognized by reviewers at URSB TISC and NCHE events as aligned with national digital education priorities \\
\bottomrule
\end{tabular}
\end{center}
\end{table}

These results, drawn from expert observation at demonstrations rather than controlled experimentation, represent practitioner-level ecological feedback from communities whose adoption and endorsement are prerequisite for meaningful educational impact. They do not substitute for quantitative efficacy measurement, which is the direct next research step.

\section{Deployment and Socioeconomic Relevance}

Arapai targets three primary deployment pathways: (i)~direct installation in school computer labs under government or donor hardware programs; (ii)~NGO-supported distribution via pre-loaded USB drives or SD cards to rural schools; and (iii)~integration into national digital-education initiatives such as Uganda's Digital Education Strategy.

From a cost perspective, Arapai eliminates recurring internet data costs (estimated at USD~2--5/student/month in Uganda), replacing them with a one-time distribution overhead. The 1.5~GB installer is small enough to distribute via SD card or USB, and the SQLite-based progress tracking can optionally sync to a central server when connectivity is intermittently available, supporting school-level analytics without hard connectivity dependencies.

Arapai aligns with OECD findings on AI's transformative potential in education \cite{b7} and the UNDP's digital transformation agenda \cite{b8}. Critically, it is designed as a \textit{complement} to teachers, not a replacement, by handling the drill, diagnosis, and formative feedback functions that are structurally infeasible at scale in under-resourced classrooms.

\section{Limitations and Future Work}

The current study has important scope limitations that define the immediate research agenda.

\textbf{Curriculum scope.} \textit{This is the most significant scope constraint of the current prototype.} The knowledge base covers only quadratic equations: one of approximately 14 O-level mathematics units, comprising six subtopics and 90 practice items. All claims in this paper regarding CBC alignment and system design generalizability are made in this context. Arapai has not been tested on topics requiring different question types (e.g., geometry, statistics, or proof-based reasoning), and no assertion of performance on those topics is warranted at this stage. Expanding the knowledge base to full O-level syllabus coverage, and validating system behavior across diverse question types, is a necessary prerequisite before broader deployment claims can be made.

\textbf{Evaluation stage.} The evidence presented in this paper is grounded in expert stakeholder observation at demonstrations rather than controlled empirical measurement. While the breadth and quality of expert consensus across nationally prominent events is substantive, it does not yield quantitative efficacy data on student learning outcomes. Controlled experimental evaluation with student cohorts is the direct next step.

\textbf{Sample and geographic scope.} Expert exposure to date, while spanning multiple Ugandan regions and institutions, does not yet constitute a systematic sampling frame. Multi-school engagement with standardized feedback protocols will provide more generalizable reception data.

\textbf{Mobile deployment.} The current implementation targets PC deployment. A mobile-optimized version for Android has not yet been validated in field conditions.

Future work will address these limitations along the following dimensions:

\textbf{Controlled efficacy study.} A pre-test/post-test randomized controlled study with O-level students at secondary schools in Uganda is planned as the immediate next research phase. The study will compare learning outcomes between students using Arapai and those receiving conventional classroom instruction across matched cohorts, with statistical analysis including effect size estimation and confidence interval reporting.

\textbf{Mobile deployment.} Gemma 2B inference has been demonstrated on mid-range Android devices (Snapdragon~778G, 8~GB RAM); a native Android port is under active development.

\textbf{Expanded multi-school evaluation.} A multi-school randomized controlled trial ($n \geq 200$) with ANOVA-based multi-group comparisons and longer follow-up periods (12 weeks) will provide definitive evidence of efficacy and generalizability.

\textbf{Teacher-facing dashboard.} A read-only progress view exportable as a PDF report will allow teachers to identify class-wide gaps surfaced by the diagnostic engine, creating a feedback loop between AI tutoring and classroom instruction.

\section{Conclusion}

This paper presented Arapai, a fully offline adaptive AI tutoring system that delivers structured, personalized mathematics instruction on commodity hardware without any internet dependency. Arapai's hybrid architecture that combines a quantized on-device language model for explanation generation with rule-based evaluation for reliability addresses the fundamental trade-off between AI fluency and deterministic correctness that has limited prior offline tutoring approaches.

Arapai has undergone substantive expert observation at four nationally significant events in Uganda, drawing reviewers from technical judging panels, university faculties, tertiary institutions, and secondary school practice. Across all contexts, observers affirmed Arapai's alignment with Uganda's CBC framework, the critical practical value of its fully offline architecture, and its potential as a deployable educational infrastructure tool.

These findings establish that Arapai is both technically credible and contextually appropriate for the problem it addresses. Controlled empirical evaluation with student populations is planned as the next phase, with the expert-validated foundation reported here providing a strong basis from which to proceed.

Offline AI is not a degraded fallback for environments without connectivity, rather it is a viable first-class educational infrastructure. As compact language models continue to improve in reasoning capability and efficiency, systems of this kind offer a scalable, low-cost mechanism for closing learning gaps in emerging markets worldwide.

\section*{Acknowledgment}

The authors thank the organizers and reviewers at the Gulu NCHE Exhibition 2026, Mbarara University Innovation Week 2026, and the Uganda Registration Services Bureau for the opportunity to present this work at the launch of the 40th Technology Support and Incubation Centre at Soroti University. The authors also thank Soroti University for institutional support throughout the development of this system.


\end{document}